\documentclass[pdftex,twocolumn,epjc3]{svjour3}          % twocolumn

\RequirePackage[T1]{fontenc}

\smartqed  % flush right qed marks, e.g. at end of proof

\RequirePackage{graphicx}
\RequirePackage{mathptmx}      % use Times fonts if available on your TeX system
\RequirePackage{flushend}
\RequirePackage[numbers,sort&compress]{natbib}
\RequirePackage[colorlinks,citecolor=blue,urlcolor=blue,linkcolor=blue]{hyperref}
\usepackage{amsmath,amssymb}
\usepackage{xspace}
\usepackage{lipsum}
\usepackage{upgreek}
\usepackage{float}
\usepackage{bbold}
\usepackage{placeins}
\usepackage{lineno}
\usepackage{microtype} % for better word spacing

\begin{document}
\title{Study of the deuterons emission time in pp collisions at the LHC via kaon-deuteron correlations}

\author{O. V\'azquez Doce\thanksref{addr1}
        \and
        D. L. Mihaylov\thanksref{addr2,addr3}
        \and
        L. Fabbietti\thanksref{addr2}
%        Second Author\thanksref{e2,addr2,addr3} %etc.
}
                    % Do not remove
%
%\offprints{}          % Insert a name or remove this line
%
\institute{INFN, Laboratori Nazionali di Frascati, Frascati, Italy\label{addr1}
          \and
Physics Department, Technische Universit\"at M\"unchen, Munich, Germany\label{addr2}      
          \and
Faculty of Physics, University of Sofia, Sofia, Bulgaria\label{addr3} 
}
\date{}
% The correct dates will be entered by Springer
%
\maketitle

\begin{abstract}
The femtoscopy correlation between positively charged kaons and deuterons measured in pp collisions at $\sqrt{s}=$ 13 TeV at the LHC with the ALICE experiment is employed to determine the upper limit for the time delay of the deuteron emission with respect to all other hadrons. Two scenarios are considered:
the first assumes that deuterons form following the decays of short-lived strong resonances, while the second assumes that deuteron production occurs simultaneously with all other primary hadrons. For both scenarios, an additional delay time can be introduced using the CECA source model and its upper limit can be extracted by fitting the femtoscopy correlation. Two models are considered for the strong K$^+$d final state interaction. For the scenario where deuterons production is affected by baryonic resonances the upper limit of the time delay is 2.25 fm/$c$, while for the primary production scenario the upper limit is 4.75 fm/$c$.
These results support the hypothesis of an early formation of the deuteron as an extended, weakly bound composite object in ultra-relativistic proton-proton collisions at the LHC.
\end{abstract}

\section{Introduction}
\label{intro}
Understanding how light nuclei form in high-energy collisions is a fundamental problem that has been explored for decades through both experimental and theoretical approaches. The key question revolves around how nuclear bound states are created and how their structure arises from the strong interaction and the principles of Quantum Chromodynamics.

The production of light nuclei has been measured in both fixed-target and collider experiments, using proton-proton (pp) and nuclear collisions up to lead-lead (Pb–Pb) across a wide range of beam energies. These measurements span from the
AGS~\cite{Bennett:1998be,Ahle:1999in,Armstrong:2000gz,Armstrong:2000gd}, to the SPS~\cite{Ambrosini:1997bf}, RHIC~\cite{Adler:2001prl,Adler:2004uy,Arsene:2010px,Agakishiev:2011ib,Adamczyk:2016gfs,Adam:2019wnb}, and the LHC~\cite{nuclei_pp_PbPb,nuclei_pp,deuteron_pp_7TeV,deuteron_pPbALICE,3He_pPb,deuteron_pp_13TeV,nuclei_pp_5TeV,nuclei_pp_13TeV_HM}, where the maximum center-of-mass energy of 13 TeV is achieved.
Theoretically, the production of light (anti)nuclei is primarily explained by statistical hadronization models or the coalescence approaches.

The thermal models are based on the hypothesis that thermal and chemical equilibrium can be achieved in relativistic nucleus-nucleus collisions \cite{PhysRevC.84.054916,PhysRevC.90.054907,Sharma:2018jqf,ThermalFist,Andronic:2017pug}. Consequently, hadron production is governed by conservation laws, the system's temperature, and baryon chemical potential. Production yields for hadrons, as well as light nuclei and hypernuclei, can be calculated employing partition functions. Fits to abundant data spanning center-of-mass energies from a few GeV to TeV \cite{Letessier:2005qe,Andronic:2017pug}, provide a satisfactory description of all measured yields. The statistical hadronization model (SHM) also successfully describes event by-event fluctuation observables, including the recently measured antideuteron number fluctuations \cite{ALICE:2022xiu}. One has though to remark that the SHM needs a different set of parameters to describe nuclei and light flavour hadrons to account for a smaller correlation volume.
Single hadron and nuclei yields can also be described in pp and p-A collisions \cite{Vovchenko:2019kes}.
 
A debated question in this field is how shallow-bound states, such as deuterons (binding energy = 2.7 MeV) or hypertritons (binding energy $\approx$150~KeV), can survive the pion-rich environment created after hadronization and be detected.
In the context of the thermal model, the hypothesis of doorway states has been proposed \cite{Braun-Munzinger:1994zkz,Andronic:2017pug}. These states are initially created as compact intermediate states with the mass and quantum numbers of the loosely bound states. They should have a lifetime of 5 fm/$c$ or longer, with an excitation energy of 40 MeV, and eventually develop into the final-state hadrons that are measured in the detector \cite{Andronic:2017pug}. Within this framework, nuclei are considered primordial, not originating from strong resonances and not influenced by them.

As an alternative to the thermal production, the assisted fusion of nucleons has been considered as a possible formation mechanism for nuclei in nucleus-nucleus collisions. These processes are studied by means of coalescence models \cite{Sun:2022xjr}. Coalescence among nucleons produced close in space and momentum can lead to the production of (anti)nuclei in high-energy collisions. Several versions of such models exist, some of which focused on the momentum correlation of the nucleons~\cite{Butler:1963,Kapusta:1980}, or employing the density matrix formalism to evaluate the formation probability of nuclei which does not contain any dynamical correlation among the nucleons \cite{Sato:1981ez}. 
Further developments of such models included detailed parameterizations of the expanding system created in heavy ion collisions \cite{Nagle:1996vp,Scheibl:1998tk} and employed a Wigner function representation of nucleons and nuclei while connecting the coalescence process to femtoscopy correlations. Femtoscopy exploits correlation in momentum space to study both the particle emitting source for the hadrons of interest and the final state interaction (Coulomb or strong) among the latter \cite{Lisa:2005dd}. 
Correlations in space are very important in the coalescence process, since the nucleons wave functions need to overlap, and have been considered exploiting event generators such as EPOS and Pythia \cite{alternativeCoalescence,KachelriessLast}.
A state-of-the-art coalescence model \cite{Mahlein:2023fmx} incorporates the latest measured femtoscopy radii for baryon-baryon pairs \cite{ALICE:2020ibs,CECA}. It utilizes the Wigner formalism to handle single nucleon densities and deuteron wave functions, and is based on event generators (EPOS and Pythia) where charged particle distributions and single hadron spectra are calibrated to experimental data. This model successfully reproduces the inclusive deuteron spectra measured in pp collisions at $\sqrt{s}$= 13 TeV by ALICE at the LHC without requiring any scaling parameters \cite{Mahlein:2023fmx}.

A more direct link between nuclei formation and final state interactions has been highlighted in several previous theoretical studies \cite{Sato:1981ez,Mrowczynski:1992gc}. However, it is the high-quality data collected in recent years by the ALICE and STAR collaborations \cite{ALICE:2023bny,STAR:2024lzt} that has opened up new opportunities in this field.
The initial idea of linking femtoscopy measurements of neutron-proton pairs to deuteron formation has not yet been realized, mainly due to the difficulty of measuring neutrons in most heavy ion experiments. However, the ALICE collaboration has conducted precise measurements of femtoscopy correlations involving deuterons paired with positively charged kaons and protons \cite{ALICE:2023bny}. These results, combined with state-of-the-art calculations \cite{Viviani:2023kxw,Torres-Rincon:2024znb}, can be used to study the evolution of deuteron states in ultra-relativistic proton-proton collisions.

In this work, the $\mathrm{K}^+$-d correlation function published in \cite{ALICE:2023bny} is employed to determine the maximal source size of the system that aligns with the measured correlations.
By utilizing the CECA source model~\cite{CECA}, the source size parameter describing the $\mathrm{K}^+$-d correlation can be translated into a time delay for the deuteron emission. This delay is defined relative to the hadronization time of all other hadrons, allowing us to determine if there is a temporal difference between this hadronization time and the formation of the deuteron that interacts with a $\mathrm{K}^+$ as observed in scattering data.
The determined upper limit for the delay time can help constraining the production mechanism for light (anti)nuclei at the LHC.

\section{Analysis}
\label{sec:Analysis}

Femtoscopy is a technique that relates particle correlations in momentum space to their emission source and interaction potential~\cite{Lisa:2005dd}. Typically, it is applied for studies of particle pairs by measuring the two-particle correlation function $C(k^*)$, where $k^*$ is the single particle momentum in the pair rest frame. The theoretical description is provided by the Koonin-Pratt relation
\begin{equation}\label{eq:KooninPratt_Simple}
C(k^*)=\int S(r^*)\left|\Psi(\vec{k}^*,\vec{r}^*)\right|^2d^3r^*,
\end{equation}
where $r^*$ is the relative distance between the pair at the time of their effective emission. The source function $S(r^*)$ represents the probability to emit a pair at a certain relative distance. The present work aims at investigating the properties of deuteron emission via the measured K$^+$d correlation function by ALICE in pp collisions at $13~\text{TeV}$~\cite{ALICE:2023bny}. This necessitates the knowledge of the wave function and in this analysis the two different descriptions as used in~\cite{ALICE:2023bny} are considered: i) using an effective range (ER) fit to the cross-section prediction at threshold anchored to the available K$^+$d scattering data~\cite{Takaki:2009ei} and ii) using the fixed-center approximation (FCA) for the well known KN interaction~\cite{Aoki:2018wug,Kamalov:2000iy}.

\subsection{Source studies}
\label{sec:Source}

 In femtoscopy studies, a standard approach to model the emission source from eq. \ref{eq:KooninPratt_Simple} is to adapt a Gaussian or a Lévy-type distribution \cite{Csorgo:2003uv}. However, recent studies of the emission source in small collision systems, such as pp collisions, have demonstrated a common source size for all hadron-hadron pairs, under the consideration of delayed particle emission due to the effect of feed-down from strongly decaying short-lived resonances \cite{ALICE:2020ibs,ALICE:2023sjd}. This idea has been practically integrated by the development of the so-called resonance source model (RSM) \cite{dimithesis}, which has been successfully used to model the emission in various types of correlation studies \cite{ALICE:2020ibs,ALICE:2023sjd,ALICE:2018ysd,ALICE:2019gcn,ALICE:2021njx,ALICE:2019buq,ALICE:2019hdt,ALICE:2020mfd,ALICE:2022uso,ALICE:2022enj,ALICE:2024bhk,ALICE:2023zbh,ALICE:2021ovd}. The RSM has a single parameter, the width of a Gaussian primordial core $r_{\text{core}}$, and uses a Monte Carlo simulation to integrate the effect of particle production through resonances. This allows the calculation of the final source distribution, which is specific to each studied pair due to differences in the relevant resonance cocktails. This distribution is often parameterized by an effective Gaussian source, with a width $r_{\text{eff}}$.

Despite the successful implementation of the RSM, the model does not allow for the inclusion of kinematic effects, such as the observed scaling of the source size as a function of the pair transverse mass ($m_\text{T}$), leading to smaller $r_{\text{eff}}$ at larger $m_\text{T}$. The origin of the $m_\text{T}$ scaling in small collision systems is still under debate, but a recent toy Monte Carlo model, called CECA \cite{CECA}, demonstrates the feasibility of generating $m_\text{T}$ scaling by introducing spatial-momentum correlations, driven by the outward expansion of the collision system with respect to the collision point. The model has been demonstrated to provide a good description of both the pp and p$\Lambda$ correlations measured by ALICE in pp collisions. In particular, both particle pairs were described using a common primordial source, and the $m_\text{T}$ scaling was reproduced. This has been achieved by using only three free fit parameters. The first parameter, $r_\text{d}$, is a random displacement parameter, on the order of $\sim 0.3~\text{fm}$, pointing to the fluctuations of the impact position for the initial hard scattering process resulting in the production of the final hadrons. The second parameter represents a hadronization scale, $h_\text{T}$, and defines an ellipsoidal surface around the collision point at which hadrons are assumed to form.
%This parameter has a typical scale of $3~\text{fm}$. 
The third parameter assumes that before the particles are free-streaming away from the source, they are propagated further for a time $\tau$.
The parameter $h_\text{T}$ combined with the small value of the displacement parameter $r_\text{d}$ introduces space-momentum correlations, as all particles propagate radially away from the collision point and this lead to the m$_T$ scaling of the source size. Typical values of $h_\text{T}$ are around $3~\text{fm}$. The parameter $\tau$ is essential for adjusting the radius size during data fitting, typically having values around $3~\text{fm}/c$.

Recent measurements of the pd and K$^+$d correlation functions by ALICE in pp collisions at $13~\text{TeV}$ \cite{ALICE:2023bny} demonstrate that both systems can be described using the $r_{\text{eff}}$ extracted from the RSM. For the pd pair, this is only possible through a dedicated three-body calculation \cite{Viviani:2023kxw}, treating the deuteron as a composite object of two nucleons. The effective radius enters the calculation to describe the separation of the three nucleons and is extracted from measurements of the pp correlation function and it is equal to r$_{eff}^{pp}= 1.43 ^{+0.16}_{-0.16}$ fm. 
The K$^+$d correlation is simpler to model due to the absence of identical nucleons, and the data is successfully described by treating it as a two-body system and employing again the $r_{\text{eff}}$ extracted from the RSM. The extracted effective source of the K$^+$d correlation is equal to r$_{eff}^{K^+d}= 1.35 ^{+0.04}_{-0.05}$ fm. 
The system considered in the $\mathrm{K}^+$-d and p-d calculations, which accounts for the Coulomb and residual strong interactions, explicitly includes the proton-neutron pair bound to a deuteron with the deuteron's quantum numbers. The extracted radii in both cases are very small and align with the universal source characterizing the production of all other hadrons. This makes it plausible to think that nuclei are produced relatively early after the collision. 
%Additionally, the fact that the calculations for the p-d case are carried out considering the effective source radius for the nucleon-nucleon pairs is compatible with the hypothesis that deuterons are formed after the creation of nucleons and following the decay of strong resonances such as the $\Delta$s.

In the present work, the published ALICE correlation function on K$^+$d is reanalyzed, by utilizing the CECA model to study the characteristic time at which the K$^+$d final state interaction signal develops. This is achieved by introducing two scenarios (A and B) to delay the deuteron emission with respect to the primordial particles. In scenario A, it is assumed that deuterons can be formed from nucleons produced from the decays of excited states, such as $\Delta$ and N$^*$ resonances. The effective mass, lifetime, and fraction of resonances decaying into nucleons in pp collisions can be extracted by the SHM \cite{Vovchenko:2019pjl}, as demonstrated by the ALICE collaboration \cite{ALICE:2023sjd}. In particular, in pp collisions at $13~\text{TeV}$, the number of primordial nucleons is $35.8\%$. 
Assuming that deuterons are formed from nucleons, by extension, the fraction of primordial deuterons will be the square ($12.8\%$) of the nucleon fraction, while the remaining $87.2\%$ include at least one nucleon produced from a resonance, leading to a delayed formation. This effect is integrated within CECA, by producing $87.2\%$ of the deuterons with an exponential delay of $1.65~\text{fm}/c$, consistent with the effective lifetime of resonances decaying into nucleons~\cite{ALICE:2020ibs}.

Within scenario B the effective emission of the deuteron is delayed by propagating all deuterons for an additional time $\tau_{\text{delay}}$, before the final state interaction with kaons sets in.
This additional delay can be applied both to primordial deuterons, as well as to deuterons formed from nucleons stemming from resonances.

\begin{figure}
% Use the relevant command for your figure-insertion program
% to insert the figure file.
% For example, with the option graphics use
\resizebox{0.48\textwidth}{!}{%
  \includegraphics{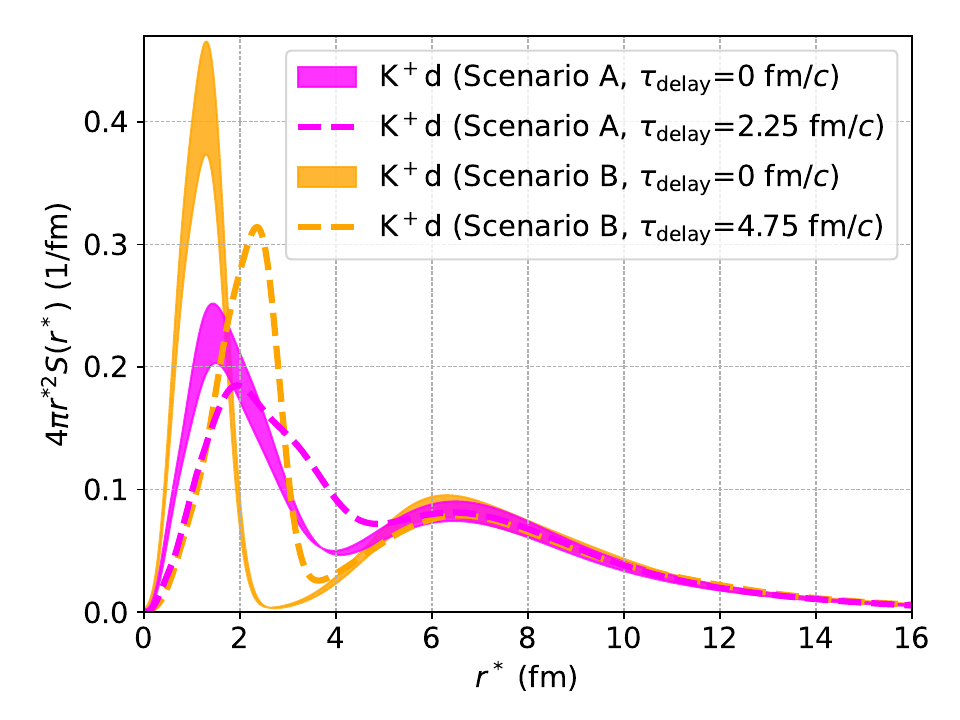}
}
% If not, use
%\vspace{5cm}       % Give the correct figure height in cm
\caption{Probability density distribution of the two-particle source for $\mathrm{K}^+$-d pairs according to the two production scenarios for deuterons. Scenario A (magenta band) considers delayed deuteron production by accounting for strong decays of baryonic resonances. Scenario B (orange band) treats all deuterons as primordial particles. The dashed lines refer to the case of additionally delaying the production of all deuterons by $\tau_\text{delay}$. See the text for details.
%Probability density distribution of the two-particle source for $\mathrm{K}^+$-d pairs according to the two scenarios of deuterons produced after the strong decay of baryonic resonances (magenta bands, scenario A) and as primordial particles (orange bands, scenario B). The solid and dashed lines refer to not-delayed and delayed emissions, respectively. See text for details.
}
\label{fig:source}       % Give a unique label
\end{figure}

Irrespective of the assumptions on deuteron emission, the kaons are modeled by considering $52.4\%$ of primordial kaons, as predicted by Thermal FIST \cite{Vovchenko:2019pjl} and \cite{ALICE:2023sjd}, while the rest stem from resonances with average lifetimes of $3.66~\text{fm}/c$. Within CECA, the K$^+$d source is evaluated after propagating the earlier-emitted particle until the time of emission of the second one. The resulting emission sources are shown in Fig.~\ref{fig:source}. The magenta band corresponds to the case of the deuterons stemming from nucleons produced from resonances. The orange band corresponds to the case of only primordial deuterons. The widths of both bands has been obtained by $10\%$ variation of the resonance contribution to the kaons. The two dotted lines show the effect of a time delay $\tau_{\text{delay}}$ of 2.25 fm/$c$ and 4.75 fm/$c$ on the deuteron emission, respectively.

All of the sources are clearly non-Gaussian and exhibit a double-peak structure. The first peak is related to the very compact core predicted by CECA, while the second structure appears due to the kaons delayed by their effective emission from a resonance. The effect is very pronounced due to the large effective lifetime and the Lorentz boost. Due to the smaller lifetime of the resonances feeding into nucleons, their inclusion translates in a broadening of the core peak. The additional delay of the deuteron (dotted lines) leads to an overall shift of all structures to larger $r^*$ values.
%
% For one-column wide figures use
%
% For two-column wide figures use
\subsection{Fits to femtoscopy data}
\label{sec:Fit}
\begin{figure*}
\resizebox{0.48\textwidth}{!}{\includegraphics{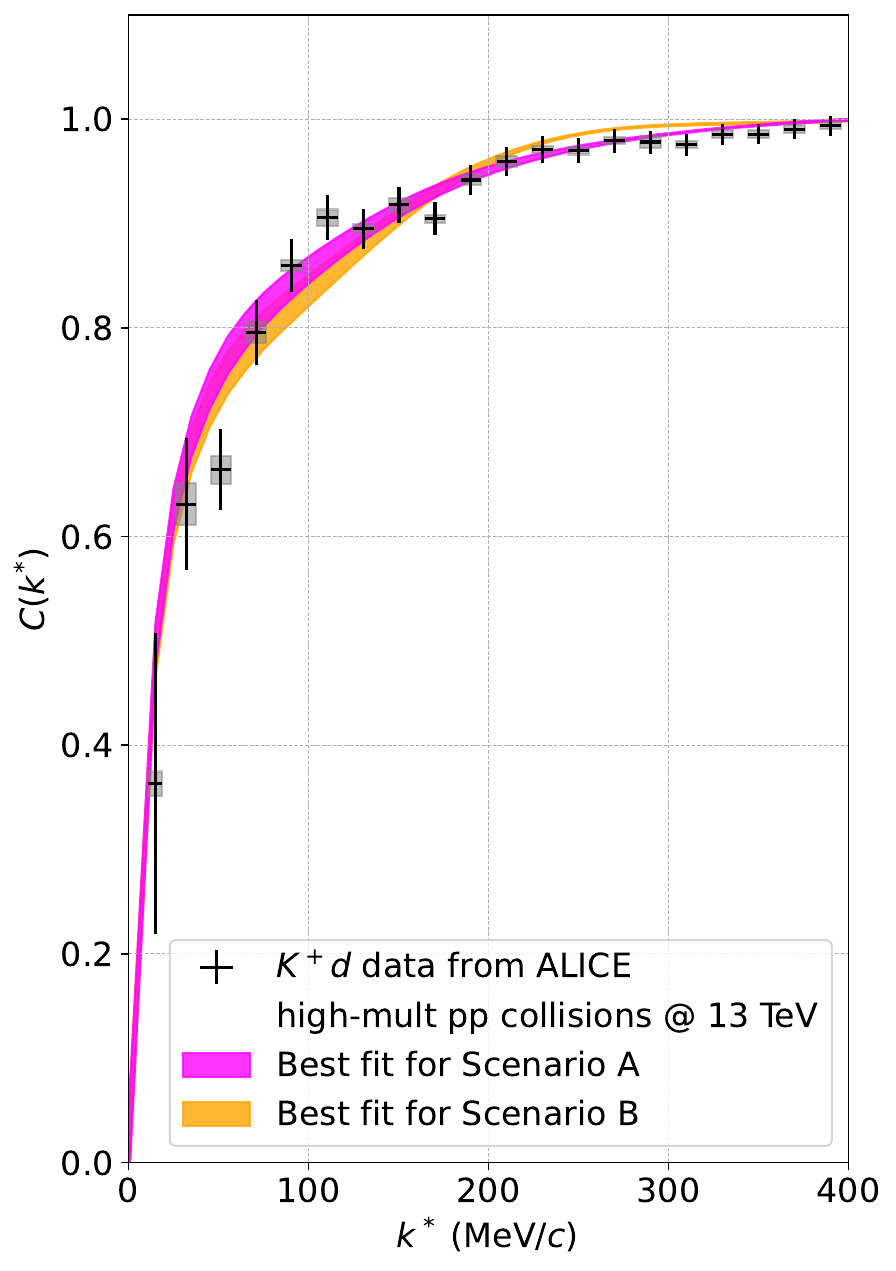}}
\resizebox{0.48\textwidth}{!}{\includegraphics{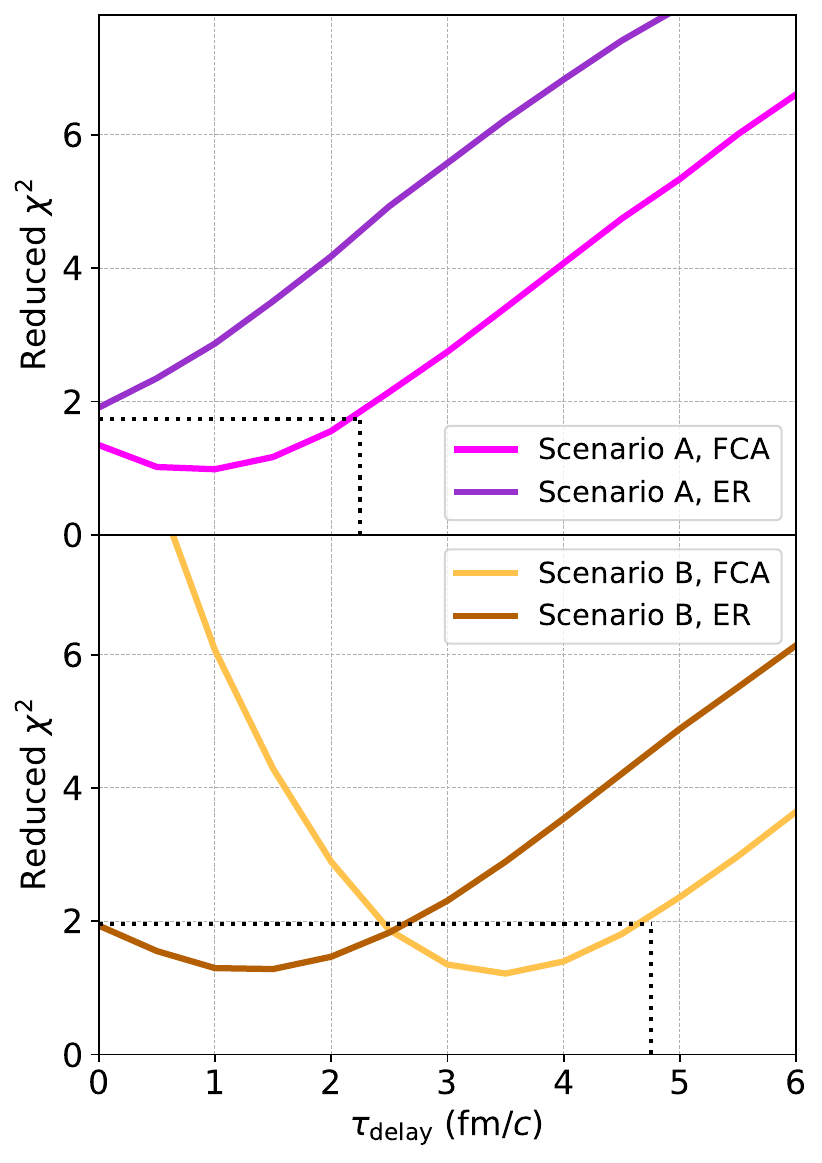}}
%\vspace{5cm}       % Give the correct figure height in cm
\caption{Left panel: Experimental K$^+$d correlation function by the ALICE Collaboration in pp collisions at $13~\text{TeV}$~\cite{ALICE:2023bny} (black points). The statistical and systematic uncertainties are shown by the black bars and gray boxes, respectively. The magenta band shows the best fit using CECA under Scenario A (deuterons emitted with a characteristic delay time due to feed-down from resonances) that corresponds to the FCA scattering parameters and a $\tau_\text{delay}$ = 1 fm/$c$. The orange band shows the best solution under Scenario B (deuterons emitted as primordial particles), that corresponds to the FCA scattering parameters and a $\tau_\text{delay}$ = 3.5 fm/$c$. Right panel: In the upper (lower) part, the reduced $\chi^2$ of the fits for Scenario A (B) are shown for both FCA and ER descriptions of the strong interaction. The dotted black lines represent the limits of the 3$\sigma$ region with respect to the best fit. See text for details.}
\label{fig:fits}       % Give a unique label
\end{figure*}

The data points in the left panel of Fig.~\ref{fig:fits} correspond to the measured K$^+$d correlation function by the ALICE Collaboration in pp collisions at $13~\text{TeV}$~\cite{ALICE:2023bny} utilizing a high-multiplicity (HM) trigger. The data is corrected for momentum resolution effects, while the non-genuine correlations related to misidentification of the selected particles and feed-down from long lived resonances are not accounted for. The ALICE Collaboration reports the fraction of genuine pairs, from correctly identified and primordial particles, to be of around 90\% (see Table III in ~\cite{ALICE:2023bny}). A flat correlation signal is associated with the non-genuine contribution. Moreover, the data is reported to contain small rising tail effect at large $k^*$, which is modeled in~\cite{ALICE:2023bny} by multiplying the expected theoretical shape by a baseline $B(k^*)=a+bk^{*2}+ck^{*3}$. Considering these effects a fit function is defined as
\begin{equation}\label{eq:CkFit}
    C_\text{fit}(k^*) = B(k^*) \left[ \lambda_\text{gen}\left(C(k^*) - 1\right) + 1\right],
\end{equation}
where for $C(k^*)$, the correlation function defined in Eq.~\ref{eq:KooninPratt_Simple} is used to fit the data, and the only free fit parameters are related to the baseline. The wave function $\Psi(\vec{k}^*,\vec{r}^*)$ is evaluated using the CATS framework~\cite{Mihaylov:2018rva} and a double Gaussian potential to model the interaction. The potential is tuned to reproduce the scattering parameters of the two interaction models introduced in chapter~\ref{sec:Analysis}. The ER and FCA approaches have scattering lengths equal to $a_0 = -0.47$~fm and $a_0 = -0.54$~fm, and effective ranges equal to $d_0 = -1.75$~fm and $d_0 = 0$~fm, respectively. The source function $S(r^*)$ is evaluated using the CECA model~\cite{CECA}, as described in section~\ref{sec:Source}. The adopted values for the parameterization of the source are taken from the best fit to the combined analysis of pp and p$\mathrm{\Lambda}$ correlations performed in~\cite{CECA}, with $r_\text{d}=0.176~$fm, $h_\text{T}=2.68~$fm and $\tau=3.76~$fm/$c$. 
The fit to the ALICE data is performed in the $k^*$ range 0-1800 MeV/c with systematic variations of $\pm$100 MeV/c allowed for the upper range. Also variations of the fraction of primordial kaons included in CECA and of the relative contribution of genuine pairs within experimental errors are allowed within 10\%.

The correlation function obtained under Scenarios A and B with this CECA source are fitted to the experimental data with the additional delay parameter $\tau_\text{delay}$ varying from 0 to 10 fm/$c$, and the value corresponding to the best fit is found for both the ER and FCA parametrisations of the K$^+$d strong interaction. In the left panel of Fig.~\ref{fig:fits} the magenta (orange) band represents the best fit under Scenario A (B), obtained for the FCA approach and a $\tau_\text{delay}$ of 3.5 (1) fm/$c$. The bandwidths reflect the allowed variations of fit range and contributions from kaon resonances in CECA and from genuine pairs in the correlation. In the upper (lower) right panel of Fig.~\ref{fig:fits} the reduced $\chi^2$ of the fits to the data are shown as magenta (orange) lines as a function of the delay $\tau_\text{delay}$ for Scenario A (B) computed in the $k^*$ range corresponding to the femtoscopic region, 0-250 MeV/c. The black dotted line shows the limits of the 3$\sigma$ region with respect to the best fit. From this region a maximum allowed $\tau_\text{delay}$ of 2.25 and 4.75 fm/$c$ is obtained for Scenario A and B, respectively. 

\section{Results}
\label{sec:2}
The results obtained from the femtoscopy fit for the upper limits of the deuteron emission time, considering two different scenarios for the deuteron production, disfavor the creation of deuterons with a time delay larger than 4.75 fm/$c$ with respect to other hadrons. Since the interaction part in the correlation function accounts for both strong and Coulomb interactions between the positively charged kaon and the nucleons forming the deuterons—either as a superposition of the K$^+$n and K$^+$p interactions or equivalent to K$^+$d scattering experiments with a deuteron target—the object studied here is not a not-interacting compact state but possesses all the properties of a fully formed deuteron.
The scenario involving the effect of the strong resonances presents lower limits for the time delay that are compatible with 0 for both interaction models, while the scenario of direct production present a lower limit of 0 and 2.5 fm/$c$ for the ER and FCA models, respectively.
The compatibility of the scenario that includes the influence of resonance decays with the experimental data, even without any additional time delay, is in agremeent with the hypothesis proposed in several coalescence approaches \cite{KachelriessLast,Mahlein:2023fmx,Sun:2022xjr} of deuteron creation via assisted nuclear fusion. This process involves nucleons originating from the decays of short-lived strong resonances.
On the other hand, it is not possible to discriminate between a direct production scenario (B) or correlated to resonance decays (A) just by evaluating the boundaries of the formation time. More stringent experimental observations and accurate modeling are necessary.

The fact that the two models for the K$^+$d interaction yield different optimal values for the delay time highlights the importance of precise interaction modeling in these studies. It also provides a method to assess systematic uncertainties in determining the upper limit of the time delay. While the results presented here for the time delay show a significant dependence on the scattering parameters used to describe the strong interaction, it is worth noting that, for the data to be compatible with delay times larger than 5 fm/$c$, an unreasonably large scattering length, one that is not compatible with current knowledge on KN interactions~\cite{Aoki:2018wug,Kamalov:2000iy} and K$^+$d scattering~\cite{Takaki:2009ei}, would be required.

The experimental data disfavor the scenario of a compact doorway state formed in ultra-relativistic pp collisions at $\sqrt{s} = 13$ TeV that reaches the correct size and binding energy after a time longer than 5 fm/$c$.

Since this analysis is conducted for high multiplicity pp collisions, the environment is comparable to that created in peripheral Pb-Pb collisions at the LHC. In peripheral Pb-Pb collisions, similar multiplicities are observed \cite{ALICE:2022xip}, making it plausible to extend these kinds of studies to such systems. This comparison highlights the potential for applying the findings from high multiplicity pp collisions to peripheral Pb-Pb collisions, thereby broadening the scope of the research.

\section{Summary}
\label{sec:3}

In this paper, the CECA source model is applied to describe the K$^+$d correlation function measured in pp collisions at $\sqrt{s} = 13$ TeV by the ALICE collaboration, as published in \cite{ALICE:2023bny}. Two different scenarios are considered: one (A) where the deuteron production is delayed according to the decay time of strong resonances, assuming they are formed via assisted nuclear fusion of all available nucleons, and another (B) where deuterons are produced simultaneously with all other hadrons.

 An additional time delay for the emission is introduced in both scenarios and the optimal value of this delay is extracted by fitting the K$^+$d correlation function. For the strong interaction part, two sets of calculations are considered. The fit results show a maximal value of 2.25 fm/$c$ and 4.75 fm/$c$ for the scenarios A and B, respectively. These small values indicate that deuterons in pp collisions at the LHC cannot be formed as compact and non interacting objects, as hypothesized in thermal models applied to heavy ion collisions.

\label{Bibliography}
\bibliographystyle{unsrtnat}
\bibliography{bibliography.bib}

\begin{thebibliography}{65}
\providecommand{\natexlab}[1]{#1}
\providecommand{\url}[1]{\texttt{#1}}
\expandafter\ifx\csname urlstyle\endcsname\relax
  \providecommand{\doi}[1]{doi: #1}\else
  \providecommand{\doi}{doi: \begingroup \urlstyle{rm}\Url}\fi

\bibitem[Bennett et~al.(1998)]{Bennett:1998be}
M.~J. Bennett et~al.
\newblock {Light nuclei production in relativistic Au + nucleus collisions}.
\newblock \emph{Phys. Rev. C}, 58:\penalty0 1155--1164, 1998.
\newblock \doi{10.1103/PhysRevC.58.1155}.

\bibitem[Ahle et~al.(1999)]{Ahle:1999in}
L.~Ahle et~al.
\newblock {Proton and deuteron production in Au + Au reactions at 11.6 A-GeV/$c$}.
\newblock \emph{Phys. Rev. C}, 60:\penalty0 064901, 1999.
\newblock \doi{10.1103/PhysRevC.60.064901}.

\bibitem[Armstrong et~al.(2000{\natexlab{a}})]{Armstrong:2000gz}
T.~A. Armstrong et~al.
\newblock {Measurements of light nuclei production in 11.5 A-GeV/$c$ Au + Pb heavy ion collisions}.
\newblock \emph{Phys. Rev. C}, 61:\penalty0 064908, 2000{\natexlab{a}}.
\newblock \doi{10.1103/PhysRevC.61.064908}.

\bibitem[Armstrong et~al.(2000{\natexlab{b}})]{Armstrong:2000gd}
T.~A. Armstrong et~al.
\newblock {Anti-deuteron yield at the AGS and coalescence implications}.
\newblock \emph{Phys. Rev. Lett.}, 85:\penalty0 2685--2688, 2000{\natexlab{b}}.
\newblock \doi{10.1103/PhysRevLett.85.2685}.

\bibitem[Ambrosini et~al.(1998)]{Ambrosini:1997bf}
G~Ambrosini et~al.
\newblock {Baryon and anti-baryon production in lead-lead collisions at 158 A-GeV/$c$}.
\newblock \emph{Phys. Lett. B}, 417:\penalty0 202--210, 1998.
\newblock \doi{10.1016/S0370-2693(97)01383-X}.

\bibitem[Adler et~al.(2001)]{Adler:2001prl}
C.~Adler et~al.
\newblock {$\overline{\mathrm{d}}$ and $^{3}\overline{\mathrm{He}}$ production in $\sqrt{s_{\rm NN}}$ = 130 GeV Au + Au collisions}.
\newblock \emph{Phys. Rev. Lett.}, 87:\penalty0 2623011--2623016, 2001.

\bibitem[Adler et~al.(2005)]{Adler:2004uy}
S.~S. Adler et~al.
\newblock {Deuteron and antideuteron production in Au + Au collisions at 200 GeV}.
\newblock \emph{Phys. Rev. Lett.}, 94:\penalty0 122302, 2005.
\newblock \doi{10.1103/PhysRevLett.94.122302}.

\bibitem[Arsene et~al.(2011)]{Arsene:2010px}
I.~Arsene et~al.
\newblock {Rapidity dependence of deuteron production in Au+Au collisions at $\sqrt{s_{\rm NN}}$ = 200 GeV}.
\newblock \emph{Phys. Rev. C}, 83:\penalty0 044906, 2011.
\newblock \doi{10.1103/PhysRevC.83.044906}.

\bibitem[Agakishiev et~al.(2011)]{Agakishiev:2011ib}
H.~Agakishiev et~al.
\newblock {Observation of the antimatter helium-4 nucleus}.
\newblock \emph{Nature}, 473:\penalty0 353, 2011.
\newblock \doi{10.1038/nature10079}.
\newblock [Erratum: Nature 475 (2011) 412].

\bibitem[Adamczyk et~al.(2016)]{Adamczyk:2016gfs}
L.~Adamczyk et~al.
\newblock {Measurement of elliptic flow of light nuclei at $\sqrt{s_{\rm NN}}=$ 200, 62.4, 39, 27, 19.6, 11.5, and 7.7 GeV at the BNL Relativistic Heavy Ion Collider}.
\newblock \emph{Phys. Rev. C}, 94\penalty0 (3):\penalty0 034908, 2016.
\newblock \doi{10.1103/PhysRevC.94.034908}.

\bibitem[Adam et~al.(2019)]{Adam:2019wnb}
Jaroslav Adam et~al.
\newblock {Beam energy dependence of (anti-)deuteron production in Au + Au collisions at the BNL Relativistic Heavy Ion Collider}.
\newblock \emph{Phys. Rev. C}, 99\penalty0 (6):\penalty0 064905, 2019.
\newblock \doi{10.1103/PhysRevC.99.064905}.

\bibitem[Adam et~al.(2016)]{nuclei_pp_PbPb}
Jaroslav Adam et~al.
\newblock {Production of light nuclei and anti-nuclei in pp and Pb-Pb collisions at energies available at the CERN Large Hadron Collider}.
\newblock \emph{Phys. Rev. C}, 93\penalty0 (2):\penalty0 024917, 2016.
\newblock \doi{10.1103/PhysRevC.93.024917}.

\bibitem[Acharya et~al.(2018)]{nuclei_pp}
Shreyasi Acharya et~al.
\newblock {Production of deuterons, tritons, $^{3}$He nuclei and their antinuclei in pp collisions at $\mathbf{\sqrt{{\textit s}}}$ = 0.9, 2.76 and 7 TeV}.
\newblock \emph{Phys. Rev.}, C97\penalty0 (2):\penalty0 024615, 2018.
\newblock \doi{10.1103/PhysRevC.97.024615}.

\bibitem[Acharya et~al.(2019{\natexlab{a}})]{deuteron_pp_7TeV}
Shreyasi Acharya et~al.
\newblock {Multiplicity dependence of (anti-)deuteron production in pp collisions at $\sqrt{s}$ = 7 TeV}.
\newblock \emph{Phys. Lett.}, B794:\penalty0 50--63, 2019{\natexlab{a}}.
\newblock \doi{10.1016/j.physletb.2019.05.028}.

\bibitem[Acharya et~al.(2020{\natexlab{a}})]{deuteron_pPbALICE}
Shreyasi Acharya et~al.
\newblock {Multiplicity dependence of light (anti-)nuclei production in p-Pb collisions at $\sqrt{s_{\rm{NN}}}$ = 5.02 TeV}.
\newblock \emph{Phys. Lett. B}, 800:\penalty0 135043, 2020{\natexlab{a}}.
\newblock \doi{10.1016/j.physletb.2019.135043}.

\bibitem[Acharya et~al.(2020{\natexlab{b}})]{3He_pPb}
Shreyasi Acharya et~al.
\newblock {Production of (anti-)$^3$He and (anti-)$^3$H in p-Pb collisions at $\sqrt{s_{\rm{NN}}}$ = 5.02 TeV}.
\newblock \emph{Phys. Rev.}, C101\penalty0 (4):\penalty0 044906, 2020{\natexlab{b}}.
\newblock \doi{10.1103/PhysRevC.101.044906}.

\bibitem[Acharya et~al.(2020{\natexlab{c}})]{deuteron_pp_13TeV}
S.~Acharya et~al.
\newblock {(Anti-)deuteron production in pp collisions at $\sqrt{s}=13 \ \text {TeV}$}.
\newblock \emph{Eur. Phys. J. C}, 80\penalty0 (9):\penalty0 889, 2020{\natexlab{c}}.
\newblock \doi{10.1140/epjc/s10052-020-8256-4}.

\bibitem[Acharya et~al.(2022{\natexlab{a}})]{nuclei_pp_5TeV}
Shreyasi Acharya et~al.
\newblock {Production of light (anti)nuclei in pp collisions at $\sqrt{s} = 5.02$~TeV}.
\newblock \emph{Eur. Phys. J. C}, 82\penalty0 (4):\penalty0 289, 2022{\natexlab{a}}.
\newblock \doi{10.1140/epjc/s10052-022-10241-z}.

\bibitem[Acharya et~al.(2022{\natexlab{b}})]{nuclei_pp_13TeV_HM}
Shreyasi Acharya et~al.
\newblock {Production of light (anti)nuclei in pp collisions at $ \sqrt{s} $ = 13 TeV}.
\newblock \emph{JHEP}, 01:\penalty0 106, 2022{\natexlab{b}}.
\newblock \doi{10.1007/JHEP01(2022)106}.

\bibitem[Cleymans et~al.(2011)Cleymans, Kabana, Kraus, Oeschler, Redlich, and Sharma]{PhysRevC.84.054916}
J.~Cleymans, S.~Kabana, I.~Kraus, H.~Oeschler, K.~Redlich, and N.~Sharma.
\newblock Antimatter production in proton-proton and heavy-ion collisions at ultrarelativistic energies.
\newblock \emph{Phys. Rev. C}, 84:\penalty0 054916, Nov 2011.
\newblock \doi{10.1103/PhysRevC.84.054916}.
\newblock URL \url{https://link.aps.org/doi/10.1103/PhysRevC.84.054916}.

\bibitem[Becattini et~al.(2014)Becattini, Bleicher, Grossi, Steinheimer, and Stock]{PhysRevC.90.054907}
Francesco Becattini, Marcus Bleicher, Eduardo Grossi, Jan Steinheimer, and Reinhard Stock.
\newblock Centrality dependence of hadronization and chemical freeze-out conditions in heavy ion collisions at $\sqrt{{s}_{NN}}=2.76$ tev.
\newblock \emph{Phys. Rev. C}, 90:\penalty0 054907, Nov 2014.
\newblock \doi{10.1103/PhysRevC.90.054907}.
\newblock URL \url{https://link.aps.org/doi/10.1103/PhysRevC.90.054907}.

\bibitem[Sharma et~al.(2019)Sharma, Cleymans, Hippolyte, and Paradza]{Sharma:2018jqf}
Natasha Sharma, Jean Cleymans, Boris Hippolyte, and Masimba Paradza.
\newblock {A Comparison of p-p, p-Pb, Pb-Pb Collisions in the Thermal Model: Multiplicity Dependence of Thermal Parameters}.
\newblock \emph{Phys. Rev. C}, 99\penalty0 (4):\penalty0 044914, 2019.
\newblock \doi{10.1103/PhysRevC.99.044914}.

\bibitem[Vovchenko and Stoecker(2019{\natexlab{a}})]{ThermalFist}
Volodymyr Vovchenko and Horst Stoecker.
\newblock Thermal-fist: A package for heavy-ion collisions and hadronic equation of state.
\newblock \emph{Comput. Phys. Commun.}, 244:\penalty0 295--310, 2019{\natexlab{a}}.
\newblock ISSN 0010-4655.
\newblock \doi{https://doi.org/10.1016/j.cpc.2019.06.024}.

\bibitem[Andronic et~al.(2018)Andronic, Braun-Munzinger, Redlich, and Stachel]{Andronic:2017pug}
Anton Andronic, Peter Braun-Munzinger, Krzysztof Redlich, and Johanna Stachel.
\newblock {Decoding the phase structure of QCD via particle production at high energy}.
\newblock \emph{Nature}, 561\penalty0 (7723):\penalty0 321--330, 2018.

\bibitem[Letessier and Rafelski(2008)]{Letessier:2005qe}
Jean Letessier and Johann Rafelski.
\newblock {Hadron production and phase changes in relativistic heavy ion collisions}.
\newblock \emph{Eur. Phys. J. A}, 35:\penalty0 221--242, 2008.

\bibitem[Acharya et~al.(2023{\natexlab{a}})]{ALICE:2022xiu}
Shreyasi Acharya et~al.
\newblock {First Measurement of Antideuteron Number Fluctuations at Energies Available at the Large Hadron Collider}.
\newblock \emph{Phys. Rev. Lett.}, 131\penalty0 (4):\penalty0 041901, 2023{\natexlab{a}}.

\bibitem[Vovchenko et~al.(2019)Vovchenko, D\"onigus, and Stoecker]{Vovchenko:2019kes}
Volodymyr Vovchenko, Benjamin D\"onigus, and Horst Stoecker.
\newblock {Canonical statistical model analysis of p-p , p -Pb, and Pb-Pb collisions at energies available at the CERN Large Hadron Collider}.
\newblock \emph{Phys. Rev. C}, 100\penalty0 (5):\penalty0 054906, 2019.
\newblock \doi{10.1103/PhysRevC.100.054906}.

\bibitem[Braun-Munzinger and Stachel(1995)]{Braun-Munzinger:1994zkz}
P.~Braun-Munzinger and J.~Stachel.
\newblock {Production of strange clusters and strange matter in nucleus-nucleus collisions at the AGS}.
\newblock \emph{J. Phys. G}, 21:\penalty0 L17--L20, 1995.
\newblock \doi{10.1088/0954-3899/21/3/002}.

\bibitem[Sun et~al.(2024)Sun, Wang, Ko, Ma, and Shen]{Sun:2022xjr}
Kai-Jia Sun, Rui Wang, Che~Ming Ko, Yu-Gang Ma, and Chun Shen.
\newblock {Unveiling the dynamics of little-bang nucleosynthesis}.
\newblock \emph{Nature Commun.}, 15\penalty0 (1):\penalty0 1074, 2024.
\newblock \doi{10.1038/s41467-024-45474-x}.

\bibitem[Butler and Pearson(1963)]{Butler:1963}
S.~T. Butler and C.~A. Pearson.
\newblock {Deuterons from High-Energy Proton Bombardment of Matter}.
\newblock \emph{Phys. Rev.}, 129:\penalty0 836--842, 1963.
\newblock \doi{10.1103/PhysRev.129.836}.

\bibitem[Kapusta(1980)]{Kapusta:1980}
Joseph~I. Kapusta.
\newblock {Mechanisms for deuteron production in relativistic nuclear collisions}.
\newblock \emph{Phys. Rev.}, C21:\penalty0 1301--1310, 1980.
\newblock \doi{10.1103/PhysRevC.21.1301}.

\bibitem[Sato and Yazaki(1981)]{Sato:1981ez}
H.~Sato and K.~Yazaki.
\newblock {On the coalescence model for high-energy nuclear reactions}.
\newblock \emph{Phys. Lett.}, B98:\penalty0 153--157, 1981.
\newblock \doi{10.1016/0370-2693(81)90976-X}.

\bibitem[Nagle et~al.(1996)Nagle, Kumar, Kusnezov, Sorge, and Mattiello]{Nagle:1996vp}
J.~L. Nagle, B.~S. Kumar, D.~Kusnezov, H.~Sorge, and R.~Mattiello.
\newblock {Coalescence of deuterons in relativistic heavy ion collisions}.
\newblock \emph{Phys. Rev.}, C53:\penalty0 367--376, 1996.
\newblock \doi{10.1103/PhysRevC.53.367}.

\bibitem[Scheibl and Heinz(1999)]{Scheibl:1998tk}
Rudiger Scheibl and Ulrich~W. Heinz.
\newblock {Coalescence and flow in ultrarelativistic heavy ion collisions}.
\newblock \emph{Phys. Rev.}, C59:\penalty0 1585--1602, 1999.
\newblock \doi{10.1103/PhysRevC.59.1585}.

\bibitem[Lisa et~al.(2005)Lisa, Pratt, Soltz, and Wiedemann]{Lisa:2005dd}
Michael~Annan Lisa, Scott Pratt, Ron Soltz, and Urs Wiedemann.
\newblock {Femtoscopy in relativistic heavy ion collisions}.
\newblock \emph{Ann. Rev. Nucl. Part. Sci.}, 55:\penalty0 357--402, 2005.
\newblock \doi{10.1146/annurev.nucl.55.090704.151533}.

\bibitem[Kachelrie{\ss} et~al.(2020)Kachelrie{\ss}, Ostapchenko, and Tjemsland]{alternativeCoalescence}
M.~Kachelrie{\ss}, S.~Ostapchenko, and J.~Tjemsland.
\newblock {Alternative coalescence model for deuteron, tritium, helium-3 and their antinuclei}.
\newblock \emph{Eur. Phys. J. A}, 56\penalty0 (1):\penalty0 4, 2020.
\newblock \doi{10.1140/epja/s10050-019-00007-9}.

\bibitem[Kachelriess et~al.(2021)Kachelriess, Ostapchenko, and Tjemsland]{KachelriessLast}
M.~Kachelriess, S.~Ostapchenko, and J.~Tjemsland.
\newblock {On nuclear coalescence in small interacting systems}.
\newblock \emph{Eur. Phys. J. A}, 57\penalty0 (5):\penalty0 167, 2021.
\newblock \doi{10.1140/epja/s10050-021-00469-w}.

\bibitem[Mahlein et~al.(2023)Mahlein, Barioglio, Bellini, Fabbietti, Pinto, Singh, and Tripathy]{Mahlein:2023fmx}
Maximilian Mahlein, Luca Barioglio, Francesca Bellini, Laura Fabbietti, Chiara Pinto, Bhawani Singh, and Sushanta Tripathy.
\newblock {A realistic coalescence model for deuteron production}.
\newblock \emph{Eur. Phys. J. C}, 83\penalty0 (9):\penalty0 804, 2023.
\newblock \doi{10.1140/epjc/s10052-023-11972-3}.

\bibitem[Acharya et~al.(2020{\natexlab{d}})]{ALICE:2020ibs}
Shreyasi Acharya et~al.
\newblock {Search for a common baryon source in high-multiplicity pp collisions at the LHC}.
\newblock \emph{Phys. Lett. B}, 811:\penalty0 135849, 2020{\natexlab{d}}.
\newblock \doi{10.1016/j.physletb.2020.135849}.

\bibitem[Mihaylov and Gonz\'alez~Gonz\'alez(2023)]{CECA}
Dimitar Mihaylov and Jaime Gonz\'alez~Gonz\'alez.
\newblock {Novel model for particle emission in small collision systems}.
\newblock \emph{Eur. Phys. J. C}, 83\penalty0 (7):\penalty0 590, 2023.
\newblock \doi{10.1140/epjc/s10052-023-11774-7}.

\bibitem[Mrowczynski(1992)]{Mrowczynski:1992gc}
S.~Mrowczynski.
\newblock {On the neutron proton correlations and deuteron production}.
\newblock \emph{Phys. Lett. B}, 277:\penalty0 43--48, 1992.

\bibitem[Acharya et~al.(2024{\natexlab{a}})]{ALICE:2023bny}
Shreyasi Acharya et~al.
\newblock {Exploring the Strong Interaction of Three-Body Systems at the LHC}.
\newblock \emph{Phys. Rev. X}, 14\penalty0 (3):\penalty0 031051, 2024{\natexlab{a}}.
\newblock \doi{10.1103/PhysRevX.14.031051}.

\bibitem[Collaboration(2024)]{STAR:2024lzt}
STAR Collaboration.
\newblock {Light Nuclei Femtoscopy and Baryon Interactions in 3 GeV Au+Au Collisions at RHIC}.
\newblock 10 2024.

\bibitem[Viviani et~al.(2023)Viviani, K\"onig, Kievsky, Marcucci, Singh, and V\'azquez~Doce]{Viviani:2023kxw}
M.~Viviani, S.~K\"onig, A.~Kievsky, L.~E. Marcucci, B.~Singh, and O.~V\'azquez~Doce.
\newblock {Role of three-body dynamics in nucleon-deuteron correlation functions}.
\newblock \emph{Phys. Rev. C}, 108\penalty0 (6):\penalty0 064002, 2023.

\bibitem[Torres-Rincon et~al.(2024)Torres-Rincon, Ramos, and Ruf\'\i{}]{Torres-Rincon:2024znb}
Juan~M. Torres-Rincon, Angels Ramos, and Joel Ruf\'\i{}.
\newblock {A two-body femtoscopy approach to the proton-deuteron correlation function}.
\newblock 10 2024.

\bibitem[Takaki(2010)]{Takaki:2009ei}
Takashi Takaki.
\newblock {Optical Potential Approach to K+ d Scattering at Low Energies}.
\newblock \emph{Phys. Rev. C}, 81:\penalty0 055204, 2010.
\newblock \doi{10.1103/PhysRevC.81.055204}.

\bibitem[Aoki and Jido(2019)]{Aoki:2018wug}
Kenji Aoki and Daisuke Jido.
\newblock {KN scattering amplitude revisited in a chiral unitary approach and a possible broad resonance in S = +1 channel}.
\newblock \emph{PTEP}, 2019\penalty0 (1):\penalty0 013D01, 2019.
\newblock \doi{10.1093/ptep/pty130}.

\bibitem[Kamalov et~al.(2001)Kamalov, Oset, and Ramos]{Kamalov:2000iy}
S.~S. Kamalov, E.~Oset, and A.~Ramos.
\newblock {Chiral unitary approach to the K- deuteron scattering length}.
\newblock \emph{Nucl. Phys. A}, 690:\penalty0 494--508, 2001.
\newblock \doi{10.1016/S0375-9474(00)00709-0}.

\bibitem[Csorgo et~al.(2004)Csorgo, Hegyi, and Zajc]{Csorgo:2003uv}
T.~Csorgo, S.~Hegyi, and W.~A. Zajc.
\newblock {Bose-Einstein correlations for Levy stable source distributions}.
\newblock \emph{Eur. Phys. J. C}, 36:\penalty0 67--78, 2004.
\newblock \doi{10.1140/epjc/s2004-01870-9}.

\bibitem[Acharya et~al.(2023{\natexlab{b}})]{ALICE:2023sjd}
Shreyasi Acharya et~al.
\newblock {Common femtoscopic hadron-emission source in pp collisions at the LHC}.
\newblock 11 2023{\natexlab{b}}.

\bibitem[Mihaylov(2021)]{dimithesis}
D.~L. Mihaylov.
\newblock {Analysis techniques for femtoscopy and correlation studies in small collision systems and their applications to the investigation of p–${\Lambda}$ and ${\Lambda}$–${\Lambda}$ interactions with ALICE}.
\newblock \emph{PhD Thesis, CERN-THESIS-2021-052}, 2021.

\bibitem[Acharya et~al.(2019{\natexlab{b}})]{ALICE:2018ysd}
Shreyasi Acharya et~al.
\newblock {p--p, p--$\Lambda$ and $\Lambda$--$\Lambda$ correlations studied via femtoscopy in pp reactions at $\sqrt{s} = 7$ TeV}.
\newblock \emph{Phys. Rev. C}, 99:\penalty0 024001, 2019{\natexlab{b}}.
\newblock \doi{10.1103/PhysRevC.99.024001}.

\bibitem[Acharya et~al.(2020{\natexlab{e}})]{ALICE:2019gcn}
Shreyasi Acharya et~al.
\newblock {Scattering studies with low-energy kaon--proton femtoscopy in proton-proton collisions at the LHC}.
\newblock \emph{Phys. Rev. Lett.}, 124:\penalty0 092301, 2020{\natexlab{e}}.
\newblock \doi{10.1103/PhysRevLett.124.092301}.

\bibitem[Acharya et~al.(2022{\natexlab{c}})]{ALICE:2021njx}
Shreyasi Acharya et~al.
\newblock {Exploring the N\ensuremath{\Lambda}\textendash{}N\ensuremath{\Sigma} coupled system with high precision correlation techniques at the LHC}.
\newblock \emph{Phys. Lett. B}, 833:\penalty0 137272, 2022{\natexlab{c}}.
\newblock \doi{10.1016/j.physletb.2022.137272}.

\bibitem[Acharya et~al.(2020{\natexlab{f}})]{ALICE:2019buq}
Shreyasi Acharya et~al.
\newblock {Investigation of the p\textendash{}\ensuremath{\Sigma^0} interaction via femtoscopy in pp collisions}.
\newblock \emph{Phys. Lett. B}, 805:\penalty0 135419, 2020{\natexlab{f}}.
\newblock \doi{10.1016/j.physletb.2020.135419}.

\bibitem[Acharya et~al.(2019{\natexlab{c}})]{ALICE:2019hdt}
Shreyasi Acharya et~al.
\newblock {First observation of an attractive interaction between a proton and a cascade baryon}.
\newblock \emph{Phys. Rev. Lett.}, 123:\penalty0 112002, 2019{\natexlab{c}}.
\newblock \doi{10.1103/PhysRevLett.123.112002}.

\bibitem[Collaboration et~al.(2020)]{ALICE:2020mfd}
Alice Collaboration et~al.
\newblock {Unveiling the strong interaction among hadrons at the LHC}.
\newblock \emph{Nature}, 588:\penalty0 232--238, 2020.
\newblock \doi{10.1038/s41586-020-3001-6}.
\newblock [Erratum: Nature 590, E13 (2021)].

\bibitem[Acharya et~al.(2023{\natexlab{c}})]{ALICE:2022uso}
Shreyasi Acharya et~al.
\newblock {First measurement of the \ensuremath{\Lambda}\textendash{}\ensuremath{\Xi} interaction in proton\textendash{}proton collisions at the LHC}.
\newblock \emph{Phys. Lett. B}, 844:\penalty0 137223, 2023{\natexlab{c}}.
\newblock \doi{10.1016/j.physletb.2022.137223}.

\bibitem[Acharya et~al.(2022{\natexlab{d}})]{ALICE:2022enj}
Shreyasi Acharya et~al.
\newblock {First study of the two-body scattering involving charm hadrons}.
\newblock \emph{Phys. Rev. D}, 106:\penalty0 052010, 2022{\natexlab{d}}.
\newblock \doi{10.1103/PhysRevD.106.052010}.

\bibitem[Acharya et~al.(2024{\natexlab{b}})]{ALICE:2024bhk}
Shreyasi Acharya et~al.
\newblock {Studying the interaction between charm and light-flavor mesons}.
\newblock \emph{Phys. Rev. D}, 110\penalty0 (3):\penalty0 032004, 2024{\natexlab{b}}.
\newblock \doi{10.1103/PhysRevD.110.032004}.

\bibitem[Acharya et~al.(2024{\natexlab{c}})]{ALICE:2023zbh}
Shreyasi Acharya et~al.
\newblock {Femtoscopic correlations of identical charged pions and kaons in pp collisions at $\sqrt{s} = 13$ TeV with event-shape selection}.
\newblock \emph{Phys. Rev. C}, 109:\penalty0 024915, 2024{\natexlab{c}}.
\newblock \doi{10.1103/PhysRevC.109.024915}.

\bibitem[Acharya et~al.(2022{\natexlab{e}})]{ALICE:2021ovd}
Shreyasi Acharya et~al.
\newblock {K$_\text{S}^0$K$_\text{S}^0$ and K$_\text{S}^0$K\ensuremath{^\pm} femtoscopy in pp collisions at $\sqrt{s} = 5.02$ and 13 TeV}.
\newblock \emph{Phys. Lett. B}, 833:\penalty0 137335, 2022{\natexlab{e}}.
\newblock \doi{10.1016/j.physletb.2022.137335}.

\bibitem[Vovchenko and Stoecker(2019{\natexlab{b}})]{Vovchenko:2019pjl}
Volodymyr Vovchenko and Horst Stoecker.
\newblock {Thermal-FIST: A package for heavy-ion collisions and hadronic equation of state}.
\newblock \emph{Comput. Phys. Commun.}, 244:\penalty0 295--310, 2019{\natexlab{b}}.
\newblock \doi{10.1016/j.cpc.2019.06.024}.

\bibitem[Mihaylov et~al.(2018)Mihaylov, Mantovani~Sarti, Arnold, Fabbietti, Hohlweger, and Mathis]{Mihaylov:2018rva}
D.~L. Mihaylov, V.~Mantovani~Sarti, O.~W. Arnold, L.~Fabbietti, B.~Hohlweger, and A.~M. Mathis.
\newblock {A femtoscopic Correlation Analysis Tool using the Schr\"odinger equation (CATS)}.
\newblock \emph{Eur. Phys. J. C}, 78:\penalty0 394, 2018.
\newblock \doi{10.1140/epjc/s10052-018-5859-0}.

\bibitem[Acharya et~al.(2023{\natexlab{d}})]{ALICE:2022xip}
Shreyasi Acharya et~al.
\newblock {Multiplicity dependence of charged-particle production in pp, p-Pb, Xe-Xe and Pb-Pb collisions at the LHC}.
\newblock \emph{Phys. Lett. B}, 845:\penalty0 138110, 2023{\natexlab{d}}.
\newblock \doi{10.1016/j.physletb.2023.138110}.

\end{thebibliography}

\end{document}